\documentclass[12pt, preprint]{revtex4-2} 
\usepackage{amsmath,amssymb} 
\usepackage{graphicx} 
\usepackage{siunitx}
\usepackage{amsmath,amssymb}
\usepackage{float} 
\usepackage{placeins}

\makeatletter
\DeclareRobustCommand{\hbar}{\mathchar'26\mkern-9mu h}
\makeatother
\newcommand{\JournalTitle}[1]{\textit{#1}}

\begin{document}

\title{Low-frequency fiber-optic vibration sensing with a Floquet-engineered optical lattice clock}

\author{Mojuan Yin$^1$, Ruohui Wang$^2$, Rui Zhou$^{2^*}$, Xueguang Qiao$^2$, Shougang Zhang $^{1,3^{*}}$}

\affiliation{Key Laboratory of Time Reference and Applications, National Time Service Center, Chinese Academy of Sciences, Xi'an, 710600, China}
\affiliation{Xi'an Key Laboratory of Optical Fiber Sensing Technology for Underground Resources, School of Physics, Northwest University, Xi'an, 710127, China}
\affiliation{School of Astronomy and Space Science, University of Chinese Academy of Sciences, Beijing, 100049, China}
\email{zhourui@nwu.edu.cn}
\email{szhang@ntsc.ac.cn}

\date{\today}

\begin{abstract}
We propose a Floquet-engineered optical lattice clock–based demodulation scheme to enhance the low-frequency performance of wound fiber-optic vibration sensors. Vibration-induced phase variations in the sensing fiber are demodulated by the Floquet-engineered Rabi spectra of the clock transition. The lattice depth with the fiber length and the Floquet-engineered Rabi spectra under the vibration from $200$ Hz down to $0.5$ Hz are simulated. With a fiber length of 4 km and transmission loss of 2 dB/km, a phase change sensitivity higher than $6 \times 10^3$ rad/g is achieved at both vibration frequencies of $200$ Hz and $0.5$ Hz.


\end{abstract}

\maketitle

\section{Introduction}

%
%

Neutral atom optical lattice clocks have achieved unprecedented frequency stability and accuracy, enabling a wide range of precision measurements beyond time and frequency metrology, including relativistic geodesy, gravitational-wave detection, and dark-matter searches~\cite{zheng2023lab, mcgrew2018atomic, mehlstaubler2018atomic, su2018low, chen2020optical}. Owing to their quantum projection noise-limited performance, optical lattice clocks can also be regarded as quantum sensors whose sensitivities surpass those of conventional optical and electronic sensors in many applications~\cite{zheng2022differential, aeppli2024clock, takano2016geopotential, safronova2018two}. Despite these advantages, operating optical lattice clocks outside well-controlled laboratory environments remains challenging. In particular, low-frequency environmental vibrations can induce lattice phase fluctuations and atomic position shifts, leading to systematic frequency shifts and degraded clock stability. Accurate and synchronous measurement of such low-frequency vibrations is therefore essential for active compensation in transportable and space-based optical clock systems.

Fiber-optic vibration sensors, especially interferometric schemes, offer attractive features for environmental monitoring, including compact size, immunity to electromagnetic interference, long-distance transmission capability, and high sensitivity~\cite{elsherif2022optical, he2021optical, wei2025large}. Coil-wound fiber configurations further enhance phase sensitivity through extended optical path lengths, making them attractive for low-frequency vibration sensing~\cite{walter2020distributed, li2021phase, li2023performance, liu2024high}. However, accurate demodulation of low-frequency and quasi-static phase signals remains challenging~\cite{gu2024ameliorated, nikitenko2018pgc}. In conventional single-frequency interferometric schemes, accumulated laser phase noise and environmental perturbations severely limit low-frequency performance, while the intrinsic 2$\pi$ phase ambiguity restricts the dynamic range and robustness.

In this work, we propose a hybrid sensing approach that combines a coil-wound fiber-optic vibration sensor with a neutral atom optical lattice clock. By integrating the fiber sensor into the lattice optical path, vibration-induced phase modulation is mapped onto atomic motion in the optical lattice and encoded in the Floquet-engineered clock transition spectra. This scheme enables high-sensitivity vibration demodulation while suppressing absolute laser phase noise as a common-mode contribution and eliminating the 2$\pi$ phase ambiguity. Furthermore, we analyze the influence of lattice-laser transmission loss in long-distance fiber links using a Floquet-theoretical model, revealing its critical role in maintaining effective demodulation performance over kilometer-scale transmission. The proposed approach provides a promising route for vibration mitigation in transportable and space-based optical clocks and offers potential applications in seismic sensing and deep-well monitoring.

\section{The vibration demodulating model based on a Floquet engineered optical lattice clock}

 In a conventional configuration, a one-dimensional optical lattice is formed by a pair of counter-propagating lattice beams along the $z$-direction, with a reflective mirror used for retro-reflection. In our scheme, the mirror is replaced by a fiber-optic vibration sensor terminated with a fiber Bragg grating (FBG), which simultaneously provides the vibration transduction and lattice laser reflection. External vibrations applied to the fiber sensor induce strain-dependent, time-varying phase modulation of the lattice light, resulting in a displacement of the optical lattice potential. Consequently, the trapped atoms following the driven lattice undergo a time-periodical motion, and the vibration information is encoded in the atomic dynamics and can be retrieved by interrogating the clock transition spectrum.

\subsection{The optical lattice potential varies with the fiber loss}
When an ultrahigh-reflectivity mirror is used to form the standing wave, the lattice-laser power loss is negligible. However, for an FBG reflector with a long transmission fiber, propagation loss significantly attenuates the reflected field, resulting in a rapid decay of the lattice depth with fiber length.
The lattice potential arises from the AC Stark shift~\cite{hofstetter2006ultracold} and is proportional to the total lattice field intensity,
  $U=-\frac{1}{2}\alpha|\vec{E}_t|^2$, where  $\alpha$ is the dipole polarizability. 
Accordingly, the longitudinal optical lattice potential along the $z$-direction can be written as
\begin{equation}
U(z,r)=-[U_{zc}+U_{z0}\cos^2(kz)]e^{-2r^2/w_0^2},
\end{equation}
The factor $e^{-2r^2/w_0^2}$ arises from transverse Gaussian intensity distribution of the lattice beams with a waist radius $w_0$ at $z=0$,where $r$ is the radial direction, $k=2\pi/\lambda_L$.  $U_{zc}$ is a constant term of the optical potential for a given incident power $P_0$, the waist radius $w_0$ and the decay rate $\kappa$. Here, $\kappa$ denotes the attenuation coefficient of the reflected electric field introduced by the fiber transmission loss and the reflectivity of the FBG.  

The one-dimensional lattice potential depth is given by $U_{z0}=4\alpha\sqrt{\kappa} P_0/(\pi c\epsilon_0w_0^2)$. In the following analysis, we consider a ${}^{87}\mathrm{Sr}$ optical lattice clock and evaluate the lattice depth $U_{z0}$ as a function of fiber length under different fiber transmission losses and FBG reflectivities. The optical power decay factor is expressed as $\kappa=R\cdot10^{-2\gamma/10}$, where $\gamma$ denotes the fiber transmission loss and $R$ is the FBG reflectivity. For the ${}^{87}\mathrm{Sr}$ optical lattice clock, a "magic-wavelength"~\cite{katori2009magic} of 813 nm is employed, at which the AC Stark shifts of the ground and excited clock states are equal, that ensures accurate frequency of the interrogated clock transition. The corresponding atomic polarizability at the "magic-wavelength" is approximately 295 a.u.. The incident lattice laser power is set to be $P_0=3\mathrm{~W}$ and it suffers from relatively high transmission loss in optical fiber; accordingly, fiber transmission losses ranging from 2 to 5 dB/km are considered for FBG reflectivities of $R$=0.9, 0.95, and 0.99 in the following analysis.
\begin{figure}[t!]
\centering
	\includegraphics[width=1\linewidth]{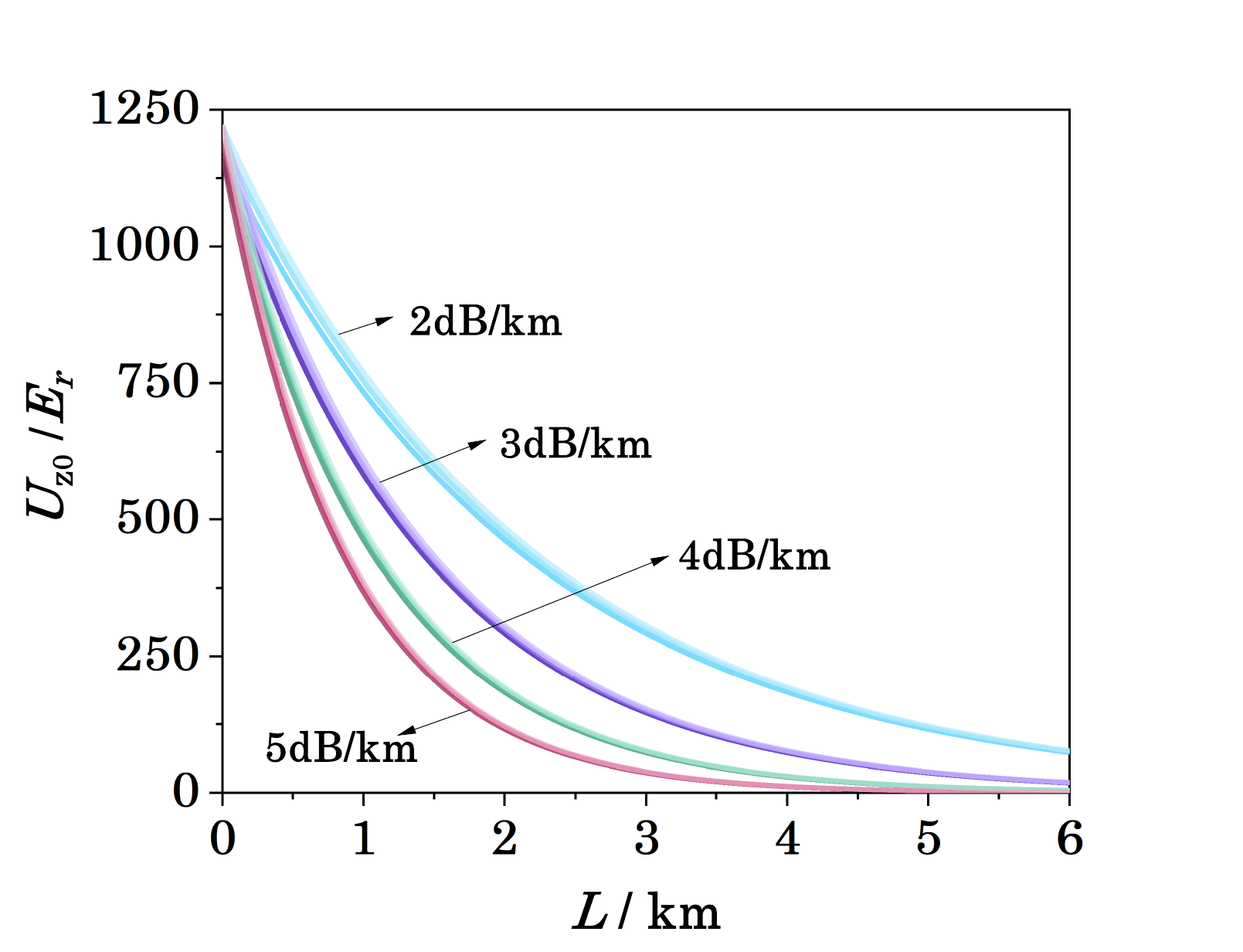}
\caption{One-dimensional optical lattice trap depth $U_{z0}$ versus fiber length. There are four groups of lines colored blue, purple, green and red, corresponding to the fiber transmission loss of 2-5 dB/km, respectively. In each group, the lightest-colored line represents the FBG reflectivity as 0.99, as the color becomes darker, the FBG reflectivity reduces to 0.95 and 0.9, respectively.}
\label{f1}
\end{figure}

Figure 1 shows the calculated longitudinal lattice potential depth $U_{z0}$, expressed in units of the recoil energy $E_r=\hbar^2 k^2/2m_a$, as a function of fiber length for different transmission losses and FBG reflectivities. The lattice depth decreases rapidly with increasing fiber length $L$ for $L$<3 km, with higher fiber loss leading to faster decay. When the transmission loss is on the order of a few dB/km, the influence of the FBG reflectivity becomes limited for $R>0.9$. For fiber lengths exceeding 4 km, the decay rate slows down, although the lattice depth remains small. 

When the lattice depth falls below approximately 20 $E_r$, tunneling between nearest-neighbor lattice sites becomes significant. In a shallow lattice, tunneling-induced spin-orbit coupling (SOC) leads to a splitting of the clock transition spectrum, known as the Van Hove splitting (VHS). The tunneling rate depends on the lattice depth $U_{z0}$, and in the lowest lattice band it decreases exponentially with increasing $U_{z0}$ (in unit of $E_r$), can be approximately written as:

\begin{equation}
\frac{J_0}{E_r}\approx\frac{4}{\sqrt{\pi}}\left(\frac{U_{z0}}{E_r}\right)^{\frac{3}{4}}\exp\left[-2\left(\frac{U_{z0}}{E_r}\right)^{1/2}\right].
\end{equation}

\subsection{A modulated optical lattice by a sinusoidal vibration signal with the fiber sensor}

When the mechanical transducer of the fiber sensor is subjected to an external vibration, the inertial mass block is excited and undergoes time-periodical motion due to the inertia. This motion induces deformation in the sensing optical fiber wound around the transducer, including variations in fiber length and changes in the effective refractive index arising from the elasto-optic effect. As a result, the optical phase of the light propagating through the sensing fiber is modulated, and the induced phase change can be expressed as:

\begin{equation}
\Delta\varphi=\frac{4\pi n_{eff}}{\lambda}C\cdot\Delta L_m,
\end{equation} 
where $\Delta L_m$ is the maximum amplitude of the fiber length variation, $n_{eff}$ is the effective refractive index, and $C$ is a dimensionless constant accounting for the elasto-optic contribution, with a typical value of approximately 0.78~\cite{kersey1997fiber}.

By considering the coupled mechanical response of the inertial mass and the wound sensing fiber~\cite{li2024sensitivity}, the maximum fiber length variation can be expressed as:
\begin{equation}
\Delta L_m=2N\frac{m_ba_v}{K_{eff}},
\end{equation}
where $N$ is the number of fiber turns, $m_b$ is the mass of the inertial block, $a_v$ is the vibration acceleration, and $K_{eff}$ denotes the effective stiffness of the spring-mass system with wound fiber. The effective stiffness determines the natural angular frequency of the mechanical transducer, $\omega_0=\sqrt{K_{eff}/m_b}$. In the low-frequency regime, where the vibration frequency is much smaller than $\omega_0$, the fiber length variation can be approximated as $\Delta L_m\approx2N {a_v/\omega_0^2}$. Accordingly, the phase-change sensitivity of the fiber-optic accelerometer is defined as:
\begin{equation}
S=\frac{\Delta\varphi}{a_v}.
\end{equation}

We propose a hybrid sensing scheme that integrates a coil-wound fiber vibration sensor into an optical lattice clock by replacing the conventional retro-reflection mirror with an FBG. A single-frequency lattice laser at the "magic-wavelength" propagates through the fiber sensor and is reflected by the FBG to form an optical lattice, enabling vibration demodulation via the clock transition response, as illustrated in Fig.~2.
\begin{figure}[H]
\centering
	\includegraphics[width=1\linewidth]{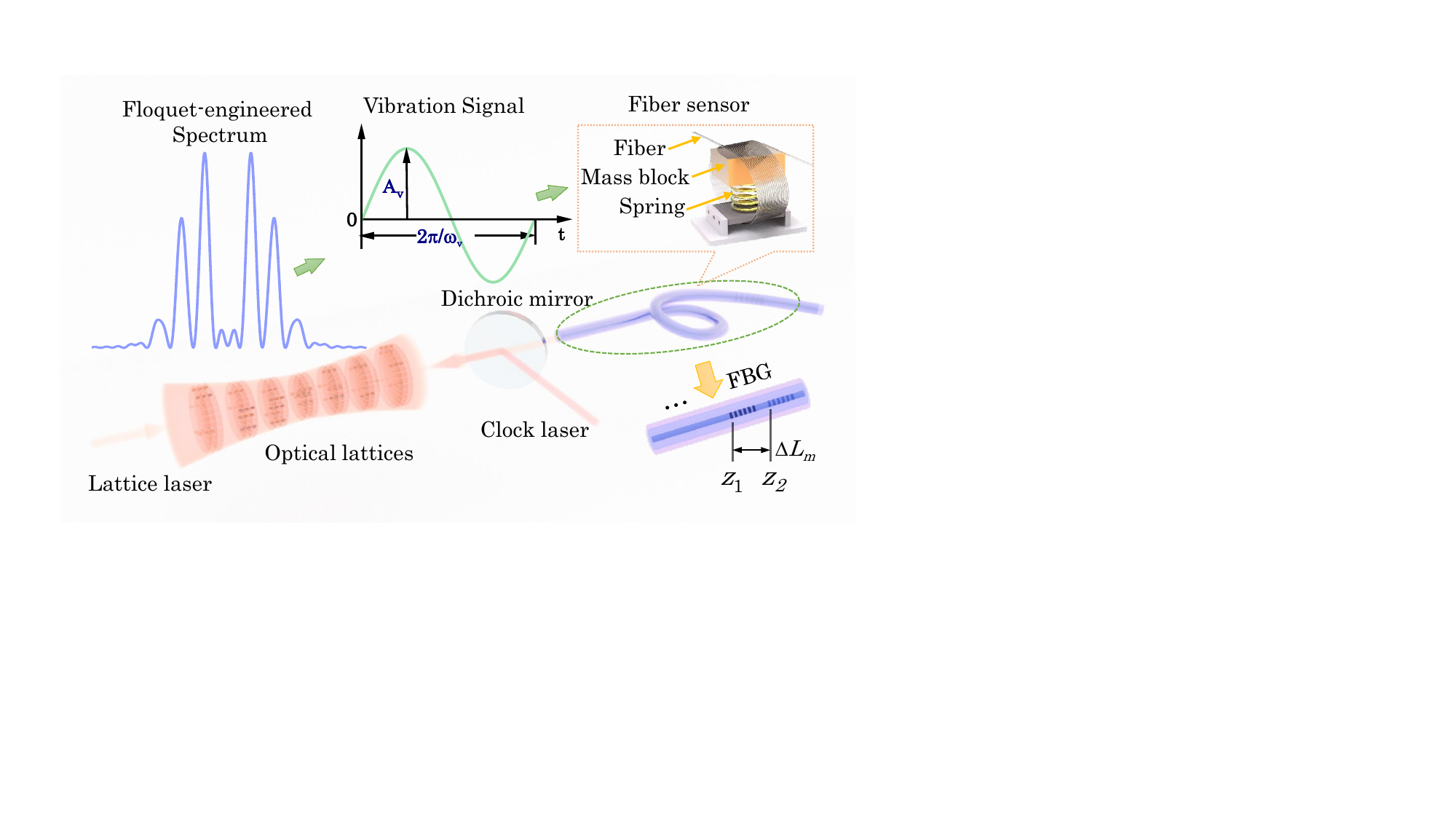}
\caption{Schematic illustration of the principle for demodulating a sinusoidal vibration signal acting on the fiber sensor by the Floquet-engineered spectrum of an optical clock. }
\label{f2}
\end{figure} 

The moving optical lattice is illustrated by three groups of periodically displaced "pancake"-shaped potential wells. The black spheres represent atoms trapped in the lattice sites. The light-red arrow denotes the lattice laser at 813 nm, while the dark-red arrow indicates the clock laser at 698 nm. The incident lattice laser propagates through the atomic ensemble and a dichroic mirror with high reflectivity at 698 nm and high transmission at 813 nm. The lattice light is then guided into the fiber, passes through the coil-wound fiber sensor, and is reflected by the FBG. External vibrations induce periodic deformation of the fiber, resulting in a time-dependent phase modulation of the lattice light. The maximum variation of fiber length $\Delta L_m$ induced by the vibration could be equivalently looked as the moving distance of the FBG.

\subsection{Demodulating the vibration by a Floquet Engineered clock transition spectrum}

We start from the theoretical model of a Floquet-engineered optical lattice clock~\cite{yin2022floquet,yin2021rabi}. The vibration signal is demodulated through the Floquet-engineered clock transition spectrum, which maps the lattice phase modulation induced by vibration onto the atomic response. We assume atoms initially occupy the lowest lattice band ($n_z=0$) and consider a one-dimensional lattice with a time-periodically modulated position~\cite{eckardt2009exploring}, given by

\begin{equation}
\widehat{H}(t)=\frac{\hat{p}^2}{2m_a}+V[\hat{z}-z_v(t)],
\end{equation}
here $z_v(t)$ describes the vibration-induced lattice displacement with angular frequency $\omega_v$. Assuming atoms remain in the lowest lattice band, the tight-binding limit allows a unitary transformation to a comoving frame, where the lattice potential is static and the vibration appears as a time-dependent inertial force acting along the lattice axis. The Hamiltonian can thus be written as:

\begin{equation}
\widehat{H}(t)=\frac{\hat{p}^2}{2m_a}+V(\hat{z})-F(t)\hat{z}.
\end{equation}

The Hubbard Hamiltonian of a driven lattice is therefore:

\begin{equation}
\widehat{H}(t)=-E_{J_0}(\widehat{T}_1+\widehat{T}_1^\dagger)+dF(t)\sum_jj|w_j\rangle\langle w_j|,
\end{equation}
here
$\hat{T}_1 $ describes nearest-neighbor tunneling with amplitude $(J_0 )$,
$\hat{z}$ is the position operator, and $j$ labels the lattice sites. The lattice constant is $ d=\lambda_L/2$, and the position operator is diagonal in the Wannier basis. The vibration-induced phase modulation is characterized by a dimensionless parameter $\zeta(t)$:

\begin{equation}
\zeta(t)=-\frac{dF(t)}{\hbar\omega_v}=\frac{md}{\hbar\omega_v}\ddot{z}_v(t).
\end{equation}

This dimensionless coefficient is defined as the ratio between the work of the inertial force $ F(t)=-m_a\ddot{z}_v(t)$ over one lattice period and the energy $\hbar\omega_v$. For a sinusoidal lattice displacement $z_v(t)=z_0sin{(\omega_vt)}$, where $z_0$ corresponds to the vibration-induced fiber length variation, the inertial force can be written as:

\begin{equation}
F(t)=-m_ad\omega_v^2n_{eff}\Delta L_m\sin{(\omega_vt)}.
\end{equation}

which leads to: 
\begin{equation}
\zeta(t)=\zeta_0\sin{(\omega_vt)},\quad\zeta_0=-m_ad\omega_vn_{eff}\Delta L_m/\hbar.
\end{equation}

The ratio between the tunneling energies with and without modulation $J_0^{\prime}/J_0 $ depends only on the modulation amplitude $\zeta_0$, and its absolute value is given by the Bessel function $ \mathcal{J}_0(\zeta_0)$. Besides the tunneling modulation, the lattice vibration also modulates the Rabi frequency $\Omega$. The unperturbed Rabi frequency is $\Omega_0=d_{eg}\mathcal{E}_0/\hbar$, and within the electric-dipole approximation the interaction Hamiltonian as $\widehat{H}_I=-\vec{d}_{eg}\cdot\vec{E}_c$. Treating the clock laser semiclassically as $ \vec{E}_c(t)=\hat{\boldsymbol{e}}_c\mathcal{E}_{c0}\mathrm{e}^{-\mathrm{i}\left(\omega_ct+\phi(t)\right)}$, the vibration-induced lattice displacement leads to a phase-modulated atom-field coupling.

For ${}^{87}\mathrm{Sr}$ atoms in a far-detuned optical lattice, the system reduces to an effective two-level model with Hamiltonian:
\begin{equation}
\widehat{H}=\widehat{H}_0+\widehat{H}_I=\hbar\begin{bmatrix}\omega_e&\Omega_0\mathrm{e}^{-\mathrm{i}\left(\omega_ct+\phi(t)\right)}\\\Omega_0\mathrm{e}^{-\mathrm{i}\left(\omega_ct+\phi(t)\right)}&\omega_g\end{bmatrix}.
\end{equation}

By shifting the zero of energy to the midpoint between the states $|g\rangle$ and $|e\rangle$, the Hamiltonian can be rewritten as:
\begin{equation}
\omega_e=-\omega_g=\frac{1}{2}(\omega_e-\omega_g)=\frac{1}{2}\omega_0.
\end{equation}

In the near-resonant regime, we define the detuning as $\delta=\omega_c-\omega_0$, and the Rabi frequency takes the form $\Omega(t)=\Omega_0\mathrm{e}^{-\mathrm{i}\left(\omega_ct+\phi(t)\right)}$. Under the rotating-wave approximation, the Hamiltonian reduces to:
\begin{equation}
\widehat{H}=\frac{\hbar}{2}\begin{bmatrix}-\delta&\Omega_0\mathrm{e}^{-\mathrm{i}\phi(t)}\\\Omega_0\mathrm{e}^{\mathrm{i}\phi(t)}&\delta\end{bmatrix}.
\end{equation}

The vibration-induced phase modulation allows the Rabi frequency to be written as 
$\Omega(t)=\Omega_0\mathrm{e}^{\mathrm{i}\phi(t)}$. 
For a sinusoidal lattice displacement, 
$\phi(t)=k_cz(t)=2k_cn_{eff}\Delta L_m\sin(\omega_vt)$, and the modulation depth is defined as
$\beta=4\pi n_{eff}\Delta L_m/\lambda_c$.
The Rabi frequency can then be expanded into a Fourier series as
 $\Omega(t)=\sum_{m=-\infty}^\infty\Omega_0\mathcal{J}_m(\beta)e^{-im\omega_vt}$,
corresponding to Floquet sidebands at $\omega_c+m\omega_v$ with effective Rabi frequencies
$ \Omega_{eff}^{(m)}=\Omega_0\mathcal{J}_m(\beta)$
and detunings
$\delta_m=\omega_c-\omega_0+m\omega_v$.
The atomic wave function can be written as:

\begin{equation}
|\psi(t^{\prime})\rangle=C_e(t^{\prime})|e\rangle+C_g(t^{\prime})|g\rangle,
\end{equation}
where $t^{\prime}$ is the clock–atom interaction time. Substituting into the Schrödinger equation and performing the transformation:
\begin{equation}
c_e^{(m)}=C_e^{(m)}e^{i\Delta t^{\prime}/2}\\c_g^{(m)}=C_g^{(m)}e^{-i\Delta t^{\prime}/2}.
\end{equation}

For $ \delta_m\neq0$, the generalized Rabi frequency of the $m$-th sideband is given by 
$\Omega_D^{(m)}=\sqrt{\Omega_0^2J_m^2(\beta)+\delta_m^2}$.
The corresponding excited-state probability amplitude after an interaction time $t^{\prime}$ can be written as:
\begin{equation}
c_e^{(m)}(t^{\prime})=\left[-i\frac{\Omega_{eff}^{(m)}}{\Omega_D^{(m)}}\sin\left(\frac{\Omega_D^{(m)}t^{\prime}}{2}\right)\right]e^{-i\Delta t^{\prime}/2}.
\end{equation}

Finally, the excited-state populations for the $m$-th sideband can be written as:
\begin{equation}
P_e^{(m)}(t^{\prime})=\left(C_e^{(m)}(t^{\prime})\right)^2=\left(\frac{\Omega_{eff}^{(m)}}{\Omega_D^{(m)}}\right)^2\sin^2\left(\frac{\Omega_D^{(m)}t^{\prime}}{2}\right).
\end{equation}

For the Floquet engineered clock transition spectrum, the excited-state population could be expressed as:

\begin{equation}
P_e(t^{\prime})=\sum_{m\in\mathbb{Z}}\left(\frac{\Omega_{eff}^{(m)}}{\Omega_D^{(m)}}\right)^2\sin^2\left(\frac{\Omega_D^{(m)}t^{\prime}}{2}\right) \label{populationE}.
\end{equation}

In the tight-binding limit, tunneling couples Bloch states in the lowest lattice band. An atom initially in $|g,q\rangle_0$ absorbs a clock-laser photon and then jump to $ |e,q+\phi\rangle_0$, the transition undergoes a momentum kick caused by the phase $\phi\approx7\pi/6$ coming from the non-uniform between wavelengths of lattice laser and clock laser. The coupling between these states forms spin-orbit coupled bands with a splitting determined by $\Omega_{eff}$ and the bandwidth of $4J_0$. For finite tunneling rate $J_0\neq0$, the modified density of states must be considered, and the clock spectrum is obtained by convolving the non-tunneling lineshape with the joint transition density of states~\cite{wall2016synthetic,kolkowitz2017spin}:

\begin{equation}
D(\delta) = 
\begin{cases} 
\frac{1}{\sqrt{[4J_0 \sin(\phi/2)]^2 - \delta^2}} & -|4J_0 \sin(\phi/2)| \leq \delta \leq |4J_0 \sin(\phi/2)| \\
0 & \text{otherwise} \label{frequency detuningD}
\end{cases}
\end{equation}

By convolving Eqs.~(\ref{populationE}) and (\ref{frequency detuningD}), a series of Floquet-engineered clock transition spectra incorporating the tunneling effect can be obtained. These spectra reveal the impact of fiber transmission loss on the vibration demodulation capability.

\section{The simulation on the vibration signal demodulating with fiber sensors}

In this work, we focus on demodulation of low-frequency vibration ranging from several hundred hertz down to sub-hertz regime. Oscillation frequencies of 200 Hz, 50 Hz, 5 Hz and 0.5 Hz are considered to simulate the Floquet engineered Rabi spectra. Figure 3 presents the calculated spectra for vibration oscillation frequencies of 200 Hz and 50 Hz with different fiber length. In these simulations, the maximum variation-induced fiber length variation is applied to the fiber sensor. The Floquet-engineered Rabi spectra are calculated for fiber lengths of 4 km and 6 km, both with a fiber loss of 4 dB/km. As fiber length increases, the SOC effect leads to progressively stronger splitting of the carrier peak of the clock transition. These spectra are interrogated by a clock laser $\pi$-pulse, which corresponding to the interrogating times of 50 ms for 200 Hz and 200 ms for 50 Hz.

For vibration frequency of 200 Hz, spectral features become unresolved when the fiber length variation $\Delta L_m$ is below 200 nm at $L$=6 km. With the vibration frequency dropping to 50 Hz, at the same condition, there is nearly no effective information about the vibration as shown in Fig. 3(d). This indicates that, for fiber length exceeding a few kilometers, the transmission loss severely limits the performance of this method in low-frequency vibration sensing. Under such high-loss conditions, vibration-induced spectral features are strongly suppressed and cannot be reliably resolved. In contrast, when the fiber transmission loss is reduced to 2 dB/km, clear demodulation features can still be observed at lower vibration frequencies as shown in Fig.4. 

\begin{figure}[htbp]
\centering
	\includegraphics[width=0.9\columnwidth]{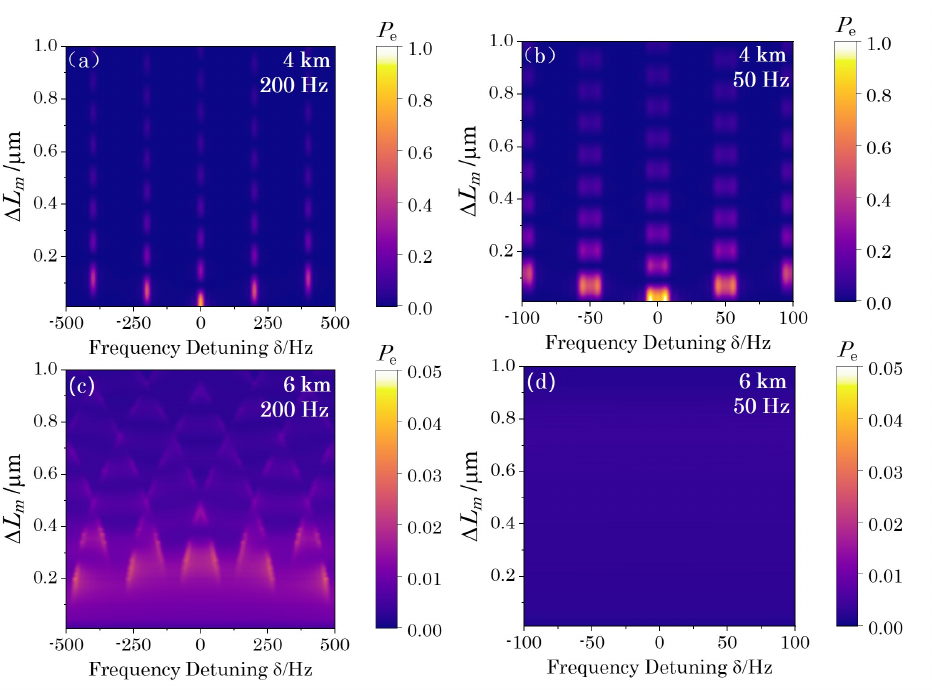}
\caption{Floquet-engineered clock transition spectra for vibration frequencies of 200 Hz and 50 Hz. The above and below panels correspond to fiber lengths of $L$=4 km and 6 km, respectively, with a fiber transmission loss of 4 dB/km.}
\label{f3}
\end{figure}

\begin{figure}[htbp]
\centering
	\includegraphics[width=0.9\columnwidth]{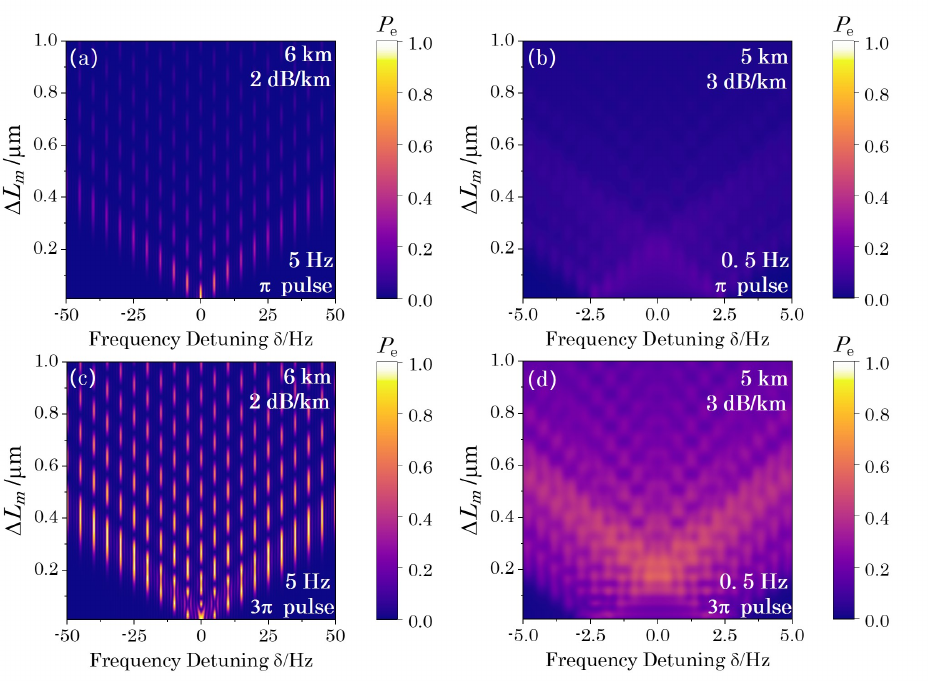}
\caption{Floquet-engineered clock transition spectra for vibration frequencies of 5 Hz and 0.5 Hz with different clock laser pulse.The spectra of above and below panels are corresponding to $\pi$-pulse and 3$\pi$-pulse,respectively.}
\label{f4}
\end{figure}

Figures 4(a) and 4(c) present the simulated Floquet engineered Rabi spectra for a vibration frequency of 5 Hz, with a fiber transmission length of 6 km and a transmission loss of 2 dB/km. Figures 4(b) and 4(d) present the simulated Floquet engineered Rabi spectra for a vibration frequency of 0.5 Hz, with a fiber transmission length of 5 km and a transmission loss of 3 dB/km. In addition, the signal-to-noise ratio (SNR) can be further improved by increasing the interrogation pulse area of the clock laser from a $\pi $-pulse to 3$\pi$-pulse. As shown in Figs. 4(c) and 4(d), the spectral intensity is significantly enhanced by employing a $\pi$-pulse compared with that in Figs. 4(a) and 4(b) by employing a 3$\pi$-pulse, indicating an improved excitation contrast and higher SNR. This enhancement arises from the increased population transfer induced by the larger pulse power, which improves their resolvability under low-frequency and long-distance transmission conditions. This provides an additional degree of freedom for optimizing the demodulation performance under practical loss conditions. This result demonstrates that reducing transmission loss is critical for extending the detectable frequency range toward the sub-100 Hz regime and highlights the importance of low-loss fiber links for long-distance, low-frequency vibration measurements using the proposed approach.

Finally, we approximately estimate the minimum detectable acceleration and the phase-change sensitivity of the proposed demodulation scheme at vibration frequencies of 200 Hz and 5 Hz. The resonance frequency of the fiber-sensor $\omega_0$ is set to be 300 Hz, and the effective refractive index is assumed to be 1.45. The number of fiber turns $N$ is set to be 58, then we could calculate the fiber length variation $\Delta L_m$ for a certain acceleration. For a fiber length of $L$=4 km with a transmission loss of 2 dB/km, the smallest detectable acceleration is first evaluated by the Floquet-engineered spectra at the vibration frequencies of 200 Hz and 5 Hz are shown in Figs. 5(a) and 5(b)，respectively.

 Assuming an atomic ensemble size on the order of $10^3$, the noise floor of the excitation fraction is typically below 0.02. As shown in Fig. 5(a), an acceleration of $\sim$ 8 $\mu$g corresponds to $\Delta L_m=2.5$ nm. When $\Delta L_m$ increases to 3 nm corresponding to an acceleration of $\sim$ 9.5 $\mu$g, the resulting increase in excitation fraction remains below this noise floor and therefore cannot be reliably resolved. In contrast, at $\Delta L_m=3.5$ nm, the excitation fraction exceeds 0.02, enabling reliable detection. Under these conditions, the minimum resolvable acceleration is estimated to be $\sim$ 3.2 $\mu$g, corresponding to a phase-change sensitivity of $6.36\times10^3$ rad/g. The discrepancy originates from the detection criterion imposed by the excitation-fraction noise floor, rather than from the absolute vibration-induced phase modulation. Similarly, at a vibration frequency of 5 Hz，the minimum detectable acceleration and resolvable acceleration are estimated to be approximately $\sim$ 24.1 $\mu$g and $\sim$ 9.4 $\mu$g, respectively, corresponding to a phase-change sensitivity of $6.5\times10^3$ rad/g. The demodulation sensitivity is limited by the atomic ensemble number, which could be increased by reducing the fiber transmission loss. And the maximum detectable acceleration is of the order of a few mg, corresponding to the fiber length $\Delta L_m$ on the order of $\mu$m. If $\Delta L_m$ is larger, atoms will escape from the optical lattice.  

\begin{figure}[H]
\centering
	\includegraphics[width=0.7\linewidth]{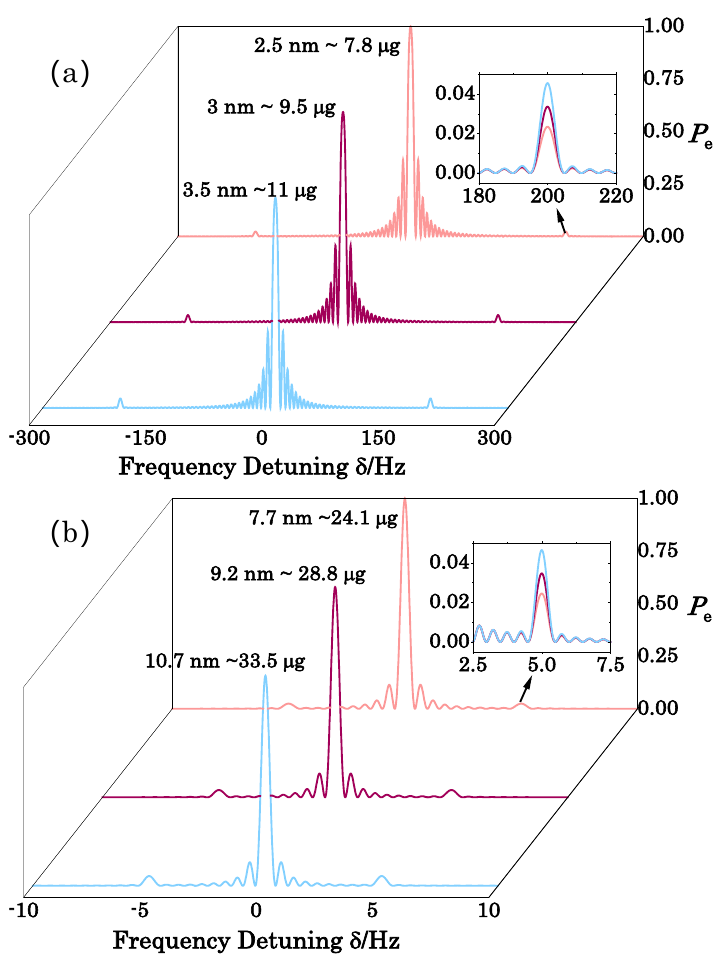}[htbp]
\caption{Floquet-engineered clock transition spectra of minimum detectable accelerations for vibration frequencies of 200 Hz and 5 Hz. }
\label{f5}
\end{figure}

\section{Conclusion}
We first established the theoretical framework of the proposed demodulation scheme by integrating a phase-interferometric coil-wound fiber sensor into the optical lattice clock system. The influence of lattice potential modulation induced by vibration-related fiber length variations was analyzed under realistic transmission losses of 2–5 dB/km at 813 nm. We then systematically investigated the Floquet-engineered Rabi spectra of the clock transition for vibration frequencies ranging from 200 Hz down to 0.5 Hz. The results show that transmission loss plays a critical role in limiting demodulation performance, particularly in the sub-100 Hz regime. With a fiber length of 4 km and a transmission loss of 2 dB/km, the phase-change sensitivity of $6.36\times 10^3$ rad/g is achieved at a vibration frequency of 200 Hz and $6.5\times 10^3$ rad/g is achieved at a vibration frequency of 5 Hz, respectively corresponding to a minimum detectable acceleration around 8 $\mu$g and 24.1 $\mu$g . Overall, the proposed method provides a viable solution for low-frequency fiber-optic vibration sensing and have potentials to improve the optical clock performance operating outside the laboratory.


\begin{thebibliography}{10}
\newcommand{\enquote}[1]{``#1''}

\bibitem{zheng2023lab}
X.~Zheng, J.~Dolde, M.~C. Cambria, \emph{et~al.}, \enquote{A lab-based test of
  the gravitational redshift with a miniature clock network,}
  {\protect\JournalTitle{Nat. Commun.}} \textbf{14}, 4886 (2023).

\bibitem{mcgrew2018atomic}
W.~F. McGrew, X.~Zhang, R.~J. Fasano, \emph{et~al.}, \enquote{Atomic clock
  performance enabling geodesy below the centimetre level,}
  {\protect\JournalTitle{Nature}} \textbf{564}, 87--90 (2018).

\bibitem{mehlstaubler2018atomic}
T.~E. Mehlst{\"a}ubler, G.~Grosche, C.~Lisdat, \emph{et~al.}, \enquote{Atomic
  clocks for geodesy,} {\protect\JournalTitle{Rep. Prog. Phys.}} \textbf{81},
  064401 (2018).

\bibitem{su2018low}
J.~Su, Q.~Wang, Q.~Wang, and P.~Jetzer, \enquote{Low-frequency gravitational
  wave detection via double optical clocks in space,}
  {\protect\JournalTitle{Class. Quantum Grav.}} \textbf{35}, 085010 (2018).

\bibitem{chen2020optical}
S.~Chen, \enquote{Optical clocks join the hunt for dark matter,}
  {\protect\JournalTitle{Physics}} \textbf{13}, s145 (2020).

\bibitem{zheng2022differential}
X.~Zheng, J.~Dolde, V.~Lochab, \emph{et~al.}, \enquote{Differential clock
  comparisons with a multiplexed optical lattice clock,}
  {\protect\JournalTitle{Nature}} \textbf{602}, 425--430 (2022).

\bibitem{aeppli2024clock}
A.~Aeppli, K.~Kim, W.~Warfield, \emph{et~al.}, \enquote{Clock with 8$\times$
  10-19 systematic uncertainty,} {\protect\JournalTitle{Phys. Rev. Lett.}}
  \textbf{133}, 023401 (2024).

\bibitem{takano2016geopotential}
T.~Takano, M.~Takamoto, I.~Ushijima, \emph{et~al.}, \enquote{Geopotential
  measurements with synchronously linked optical lattice clocks,}
  {\protect\JournalTitle{Nat. Photonics}} \textbf{10}, 662--666 (2016).

\bibitem{safronova2018two}
M.~S. Safronova, S.~G. Porsev, C.~Sanner, and J.~Ye, \enquote{Two clock
  transitions in neutral yb for the highest sensitivity to variations of the
  fine-structure constant,} {\protect\JournalTitle{Phys. Rev. Lett.}}
  \textbf{120}, 173001 (2018).

\bibitem{elsherif2022optical}
M.~Elsherif, A.~E. Salih, M.~G. Mu{\~n}oz, \emph{et~al.}, \enquote{Optical
  fiber sensors: Working principle, applications, and limitations,}
  {\protect\JournalTitle{Adv. Photonics Res.}} \textbf{3}, 2100371 (2022).

\bibitem{he2021optical}
Z.~He and Q.~Liu, \enquote{Optical fiber distributed acoustic sensors: A
  review,} {\protect\JournalTitle{J. Lightwave Technol.}} \textbf{39},
  3671--3686 (2021).

\bibitem{wei2025large}
Y.~Wei, J.~Yang, C.~Liu, \emph{et~al.}, \enquote{Large detection range and high
  sensitivity fiber optic interferometric micro-displacement sensor based on an
  n-type structure,} {\protect\JournalTitle{Opt. Express}} \textbf{33},
  35852--35864 (2025).

\bibitem{walter2020distributed}
F.~Walter, D.~Gr{\"a}ff, F.~Lindner, \emph{et~al.}, \enquote{Distributed
  acoustic sensing of microseismic sources and wave propagation in glaciated
  terrain,} {\protect\JournalTitle{Nat. Commun.}} \textbf{11}, 2436 (2020).

\bibitem{li2021phase}
Y.~Li, Y.~Wang, L.~Xiao, \emph{et~al.}, \enquote{Phase demodulation methods for
  optical fiber vibration sensing system: A review,}
  {\protect\JournalTitle{IEEE Sens. J.}} \textbf{22}, 1842--1866 (2021).

\bibitem{li2023performance}
M.~Li, X.~Li, D.~Xu, and H.~Li, \enquote{Performance analysis of the fiber
  coils combining hybrid polarization-maintaining fiber designs and symmetrical
  winding patterns,} {\protect\JournalTitle{Opt. Express}} \textbf{31},
  22424--22443 (2023).

\bibitem{liu2024high}
C.~Liu, C.~Guo, Y.~Shao, \emph{et~al.}, \enquote{High sensitivity strain and
  refractive index mzi-spr sensing based on dual fiber winding structure,}
  {\protect\JournalTitle{Opt. Express}} \textbf{32}, 24293--24303 (2024).

\bibitem{gu2024ameliorated}
S.~Gu, G.~Zhang, Q.~Ge, \emph{et~al.}, \enquote{Ameliorated 3$\times$ 3
  coupler-based demodulation algorithm using iteratively reweighted ellipse
  specific fitting,} {\protect\JournalTitle{Opt. Express}} \textbf{32},
  1108--1122 (2024).

\bibitem{nikitenko2018pgc}
A.~N. Nikitenko, M.~Y. Plotnikov, A.~V. Volkov, \emph{et~al.},
  \enquote{Pgc-atan demodulation scheme with the carrier phase delay
  compensation for fiber-optic interferometric sensors,}
  {\protect\JournalTitle{IEEE Sens. J.}} \textbf{18}, 1985--1992 (2018).

\bibitem{hofstetter2006ultracold}
W.~Hofstetter, \enquote{Ultracold atoms in optical lattices: tunable quantum
  many-body systems,} {\protect\JournalTitle{Philos. Mag.}} \textbf{86},
  1891--1906 (2006).

\bibitem{katori2009magic}
H.~Katori, K.~Hashiguchi, E.~Y. Il’inova, and V.~Ovsiannikov, \enquote{Magic
  wavelength to make optical lattice clocks insensitive to atomic motion,}
  {\protect\JournalTitle{Phys. Rev. Lett.}} \textbf{103}, 153004 (2009).

\bibitem{kersey1997fiber}
A.~D. Kersey, M.~A. Davis, H.~J. Patrick, \emph{et~al.}, \enquote{Fiber grating
  sensors,} {\protect\JournalTitle{J. Lightwave Technol.}} \textbf{15},
  1442--1463 (1997).

\bibitem{li2024sensitivity}
X.~Li, F.~Geng, S.~Li, \emph{et~al.}, \enquote{Sensitivity enhancement of
  fiber-optic accelerometers based on the weak fiber bragg grating array by
  inscribing in thin-cladding fiber,} {\protect\JournalTitle{J. Lightwave
  Technol.}}  (2024).

\bibitem{yin2022floquet}
M.-J. Yin, X.-T. Lu, T.~Li, \emph{et~al.}, \enquote{Floquet engineering
  hz-level rabi spectra in shallow optical lattice clock,}
  {\protect\JournalTitle{Phys. Rev. Lett.}} \textbf{128}, 073603 (2022).

\bibitem{yin2021rabi}
M.-J. Yin, T.~Wang, X.-T. Lu, \emph{et~al.}, \enquote{Rabi spectroscopy and
  sensitivity of a floquet engineered optical lattice clock,}
  {\protect\JournalTitle{Chin. Phys. Lett.}} \textbf{38}, 073201 (2021).

\bibitem{eckardt2009exploring}
A.~Eckardt, M.~Holthaus, H.~Lignier, \emph{et~al.}, \enquote{Exploring dynamic
  localization with a bose-einstein condensate,} {\protect\JournalTitle{Phys.
  Rev. A}} \textbf{79}, 013611 (2009).

\bibitem{wall2016synthetic}
M.~L. Wall, A.~P. Koller, S.~Li, \emph{et~al.}, \enquote{Synthetic spin-orbit
  coupling in an optical lattice clock,} {\protect\JournalTitle{Phys. Rev.
  Lett.}} \textbf{116}, 035301 (2016).

\bibitem{kolkowitz2017spin}
S.~Kolkowitz, S.~Bromley, T.~Bothwell, \emph{et~al.},
  \enquote{Spin--orbit-coupled fermions in an optical lattice clock,}
  {\protect\JournalTitle{Nature}} \textbf{542}, 66--70 (2017).

\end{thebibliography}
\end{document}